# An Enhanced Time Space Priority Scheme to Manage QoS for Multimedia Flows transmitted to an end user in HSDPA Network

Mohamed HANINI [1,4], Abdelali EL BOUCHTI[1,4], Abdelkrim HAQIQ[1,4], Amine BERQIA[2,3,4]

1 Computer, Networks, Mobility and Modeling laboratory
Department of Mathematics and Computer
FST, Hassan 1st University, Settat, Morocco
2 ENSIAS, Mohammed V Souissi University, Rabat, Morocco
3 Universiy Algarve, LG, Portugal
4 e-NGN Research group, Africa and Middle East

E-mails: {haninimohamed, a.elbouchti, ahaqiq, berqia}@gmail.com

*Abstract*— **When different type of packets with different needs of Quality of Service (QoS) requirements share the same network resources, it became important to use queue management and scheduling schemes in order to maintain perceived quality at the end users at an acceptable level. Many schemes have been studied in the literature, these schemes use time priority (to maintain QoS for Real Time (RT) packets) and/or space priority (to maintain QoS for Non Real Time (NRT) packets). In this paper, we study and show the drawback of a combined time and space priority (TSP) scheme used to manage QoS for RT and NRT packets intended for an end user in High Speed Downlink Packet Access (HSDPA) cell, and we propose an enhanced scheme (Enhanced Basic-TSP scheme) to improve QoS relatively to the RT packets, and to exploit efficiently the network resources. A mathematical model for the EB-TSP scheme is done, and numerical results show the positive impact of this scheme.**

I. INTRODUCTION

In recent years, the performance of mobile cellular telecommunication networks have been growing continuously by increasing the hardware capacity, and new generation of mobile networks offer more bandwidth resources. With this development, new services with high bandwidth demand and different QoS requirements have been incorporated and its effect needs to be taken in consideration.
Despite of the efforts taken on the infrastructures to improve network services, the disturbing impact of the wireless transmission may lead to a degradation of the perceived quality at the end users. It becomes important to take additional measures on the networks.
Hence, two ways are possible. The first is to adapt the contenent to the current network conditions at the end user. This is the end to end QoS control [15]. The most well known mechanisms to achieve this adaptation are Random Early Detection (RED) [8] and its variants [7]. The second way is to manage network resources to offer network support for content; it is a network centric approach. One of the most important representatives of this second way is queue management and packet scheduling which have impact on the QoS attributes. When different type of packets with different needs of QoS standards share the same network resources, such as buffers and bandwidth, a priority scheme from the second way has to be used. The priority scheme can be defined in terms of a policy determining [13]**:**
- Which of the arriving packets are admitted to the buffer and how it is admitted
  And/or
- Which of the admitted packets is served next

The former priority service schemes referred to as space priority schemes and attempt to minimize the packet loss of non real time (NRT) applications (www browsing, e-mail, ftp, or data access) for which the loss ratio is the restrictive quantity. The latter priority service schemes are referred as time priority schemes and attempt to guarantee acceptable delay boundaries to real time (RT) applications (voice or video) for which it is important that delay is bounded.
Many priority schemes have been studied in literature, and have focused on space priority or time priority.
Authors in [14] present a modeling for a multimedia traffic in a shared channel, but they take in consideration system details rather the characteristics of the flows composing the traffic.
Works in [1], [4], [12] study priority schemes and try to maximize the QoS level for the RT packets, without taking into account the effect on degradation of the QoS for NRT packets.
In HSDPA (High-Speed Downlink Packet Access) technology, it is possible to implement Packet scheduling algorithms that support multimedia traffic with diverse concurrent classes of flows being transmitted to the same end user [9]. Therefore, Suleiman and all present in [16] a queuing model for multimedia traffic over HSDPA channel using a combined time priority and space priority (TSP priority) with threshold to control QoS measures of the both RT and NRT packets.

The basic idea of TSP priority [2] is that, in the buffer, RT packets are given transmission priority (time priority), but the number accepted of this kind of packets is limited. Thus, TSP scheme aims to provide both delay and loss differentiation.

Authors in [16], [17] studied an extension of TSP scheme incorporating thresholds to control the arrival packets of NRT packets (Active TSP scheme), and show, via simulation (using OPNET), that TSP scheme achieves better QoS measures for both RT and NRT packets compared to FCFS (First Come First Serve) queuing.

To model the TSP scheme, mathematical tools have been used in [18] and QoS measures have been analytically deducted, but some given results are false, ([5],[6],[9]) corrected this paper and used MMPP and BMAP processes to model the traffic sources.

When the basic TSP scheme is applied to a buffer in Node B (in HSDPA technology) arriving RT packets will be queued in front of the NRT packets to receive priority transmission on the shared channel. A NRT packet will be only transmitted when no RT packets are present in the buffer, this may the RT QoS delay requirements would not be compromised [2].

In order to fulfil the QoS of the loss sensitive NRT packets, the number of admitted RT packets, is limited to R, to devote more space to the NRT flow in the buffer.

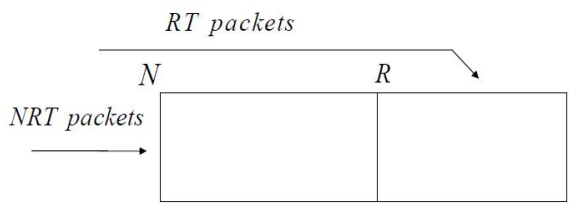

*Fig.1. the B-TSP scheme applied to a buffer*

This scheme has in important drawback; as the number of NRT packets can not exceed a threshold R, this will result in RT packet drops even when capacity is available in the section reserved to NRT packets in the buffer that implies bad QoS management for RT packets, and bad management for buffer space.

Hence, in this paper, we propose an algorithm to enhance the basic TSP scheme (Enhanced Basic TSP: EB-TSP). The priority function is modified for packets to overcome the drawback cited above, in order to improve QoS for RT packet by reducing the loss probability of RT packets, and to achieve a better management for the network resources.

The rest of this paper is organized as follows: section 2 introduces the proposed buffer management scheme, which is termed as *EB-TSP* vs. *Basic-TSP*. Subsequently, in section 3 the mathematical model is presented and studied. The QoS measures related to the proposed scheme are analytically presented in section 4. Section 5 presents the numerical results and shows the effect that the proposed scheme has on the performance of traffic. Finally, section 6 provides the concluding remarks.

II. EB-TSP SCHEME DESCRITION

The Basic-TSP (B-TSP) buffer management scheme for multimedia QoS control in HSDPA Node B, proposed by authors in [3] is defined to maintain inter-class prioritization for end-users with multiple flows. It consists on putting a buffer, for each user, where RT and NRT flows are queued according to the following scheme priority.

The RT flow packets are queued ahead of the NRT flow packets of the same user, for priority scheduling/transmission on the shared channel (time priority). At the same time, the NRT flow packets get space priority in the user's buffer queue. B-TSP scheme queuing uses a threshold $R$ to restrict the maximum number of queued RT packets (fig.1).

In [18] authors have shown B-TSP to be an effective queuing mechanism for joint RT and NRT QoS compared to conventional priority queuing schemes.

To overcome the drawback of B-TSP scheme cited in section I, we propose to use the following control mechanism:

When an RT packet arrives at the buffer, either it is full or there is free space. In the first case, if the number of RT packets is less than $R$, then an NRT packet will be rejected and the arriving RT packet will enter in the buffer. Or else, the arriving RT packet will be rejected. In the second case, the arriving RT packet will enter in the buffer.

The same, when an NRT packet arrives at the buffer, either it is full or there is free space. In the first case, if the number of RT packets is less than $R$, then the arriving NRT packet will be rejected. Or else, an RT packet will be rejected and the arriving NRT packet will enter in the buffer. In the second case, the arriving NRT packet will enter in the buffer.

**Remark:** In the buffer, the RT packets are placed all the time in front of the NRT packets.

III. MATHEMATICAL MODEL

*A. Arrival and Sevice Processes*

The arrival processes of RT and NRT packets are assumed to be poissonian with rates $\lambda_{RT}$ and $\lambda_{NRT}$ respectively.

The service times of RT and NRT packets are assumed to be exponential with rate $\mu_{RT}$ and $\mu_{NRT}$ respectively.

We also assume that the arrival processes and the service times are mutually independent between them.

The state of the system at any time t can be described by the process $X(t) = (X_1(t), X_2(t))$, where $X_1(t)$ (respectively $X_2(t)$) is the number of RT (respectively of NRT) packets in the buffer at time t..

The state space of $X(t)$ is $E=\{0,...., N\}x\{0,...., N\}$.

*B. Stability*

Since the arrival processes are Poisson (i.e the inter-arrivals are exponential), the service times are exponential and these processes are mutually independent between them, then *X(t)* is a Markov process.

We can prove easily that *X(t)* is irreducible, because all the states communicate between them.

Moreover, E is a finite space, then *X(t)* is positive recurrent. Consequently, *X(t)* is an ergodic process and the equilibrium probability exists.

*C. Equilibrium Probability*

We denote the equilibrium probability of *X(t)* at the state *(i,j)* by $\{p(i,j)\}$, where:

$$p(i,j) = \lim_{t \to \infty} P(X_1(t) = i, X_2(t) = j)$$

It is the solution of the following balance equations:

$$(\lambda_{NRT} + \lambda_{RT}) p(0,0) = \mu_{NRT} p(0,1) + \mu_{RT} p(1,0)$$

$$(\lambda_{RT} + \mu_{NRT}) p(0,N) = \lambda_{NRT} p_2(0, N-1)$$

$$(\lambda_{NRT} + \mu) p(N,0) = \lambda_{RT} p(N-1,0)$$

For *i =1, ......, N-1*

$$(\lambda_{NRT} + \mu_{RT} + \lambda_{RT}) p(i,0) = \lambda_{RT} p(i-1,0) + \mu_{RT} p(i+1,0)$$

For *j=1, ....., N-1*

$$(\lambda_{RT} + \lambda_{RT} + \mu_{NRT}) p(0,j) = \mu_{RT} p(1,j) + \lambda_{NRT} p(0,j-1) + \mu_{NRT} p(0,j+1)$$

For *i= R+1,....., N-1*

$$(\mu_{RT} + \lambda_{NRT}) p(i, N-i) = \lambda_{RT} p(i, N-i-1) + \mu_{RT} p(i-1, N-i)$$

For *i =1, ......., N-1*

$$(\mu_{RT} + \lambda_{RT}) p(i, N-i) = +\lambda_{NRT} p(i, N-i-1) + \lambda_{RT} p(i-1, N-i)$$

For *i =1, ......, N-2, j=1,...., N-i-1*

$$(\lambda_{NRT} + \mu_{RT} + \lambda_{RT}) p(i,j) = \lambda_{RT} p(i-1,j) + \lambda_{NRT} p(i,j-1) + \mu_{RT} p(i+1,j)$$

The equilibrium probability must verify the normalization equation given by: $\sum_{i=0}^{N} \sum_{j=0}^{N-i} p(i,j) = 1$.

IV. QOS MEASURES

In this section, the loss probability and the delay for each class of traffic are analytically presented.

*A. Loss Probability*

With the EB-TSP scheme, an RT packet is lost either when the buffer is full and the number of RT packets is more than *R* at the time of its arrival or when an NRT packet arrives and finds the buffer full and the number of RT packets is more than *R*.

Then the loss probability of RT packets is given by:

$$P_{L-RT} = \lim_{t \to \infty} \frac{\int_0^t 1_{(X_1(s) + X_2(s) = N, X_1(s) \geq R)}(s) A^1(s) ds}{N_1(t)} +$$

$$\lim_{t \to \infty} \frac{\int_0^t 1_{(X_1(s) + X_2(s) = N, X_1(s) \succ R)}(s) A^2(s) ds}{N_1(t)}$$

Where:

$N_1(t)$ is the number of arriving RT packets in the buffer during the time interval [0,t]

$A^1(s)$ (respectively $A^2(s)$) is the RT (respectively NRT) arriving flow in the buffer at time *s*.

$$1_{(s)}(t) = \begin{cases} 1 & \text{if } s = t \\ 0 & \text{else} \end{cases}$$

Since *X* is ergodic, we show that:

$$P_{L-RT} = \sum_{i=R}^{N} p(i, N-i) + \frac{\lambda_{NRT}}{\lambda_{RT}} \sum_{i=R+1}^{N} p(i, N-i)$$

Using the same analysis, we can show that the loss probability of NRT packets is:

$$P_{L-NRT} = \sum_{i=0}^{R} p(i, N-i) + \frac{\lambda_{RT}}{\lambda_{NRT}} \sum_{i=0}^{R-1} p(i, N-i)$$

*B. Average Number of Packets in the Buffer*

The average number of RT packets in the buffer at the steady state is:

$$N_{RT} = \lim_{t \to \infty} \frac{N_1(t)}{t}$$

We can show that:

$$N_{RT} = \sum_{i=0}^{N} \sum_{j=0}^{N-i} p(i,j)$$

We show also that the average number of NRT packets in the buffer at the steady state is:

$$N_{NRT} = \sum_{j=0}^{N} \sum_{i=0}^{N-j} p(i,j)$$

*C. Mean Delay*

Using Little's Formula [10], we deduct that the average delays of RT and NRT packets respectively are given:

$$D_{RT} = \frac{N_{RT}}{\lambda_{RT}(1 - P_{L-RT})}$$

$$D_{NRT} = \frac{N_{RT} + N_{NRT}}{\lambda_{NRT}(1 - P_{L-NRT})}$$

## V. NUMERICAL RESULTS

In this section we present the numerical results of EB-TSP scheme. We use the Maple software to solve numerically the system of equations given in III-C and to evaluate the QoS measures. The numerical results for the EB-TSP scheme are compared to the same value for basic-TSP scheme. In the simulations, we use the following parameters:

| Total queue length | 60 |
| Threshold for number of RT packets | 15 |
| Arrival rate of NRT packets | 8 |
| Rate service of RT packets | 30 |
| Rate service of NRT packets | 25 |

*Table 1 : Simulation parameters*

Figure.2 plots the loss probability for the RT packets in both B-TSP and EB-TSP schemes. This figure shows that the proposed scheme has a significant impact on the performance of the system relatively to the RT packet loss, this effect is more important when the arrival rate of RT packets is growing. Which leads to the better quality for audio and video calls received by the end user in HSDPA cell using EB-TSP scheme.

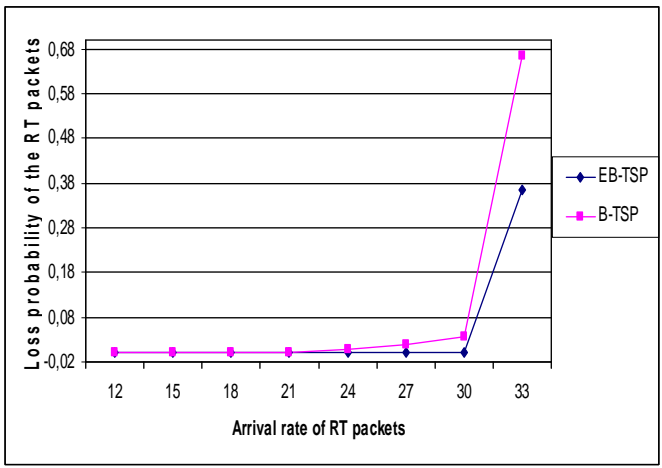

*Figure2: Variation of the loss probability of RT packets according to arrival rate of RT packets*

As expected, Figures 3, 4 and 5 show that EB-TSP scheme keeps the same level of other QoS measures: dropping probability for NRT packets and average delays for RT and NRT packets, compared to basic-TSP scheme.

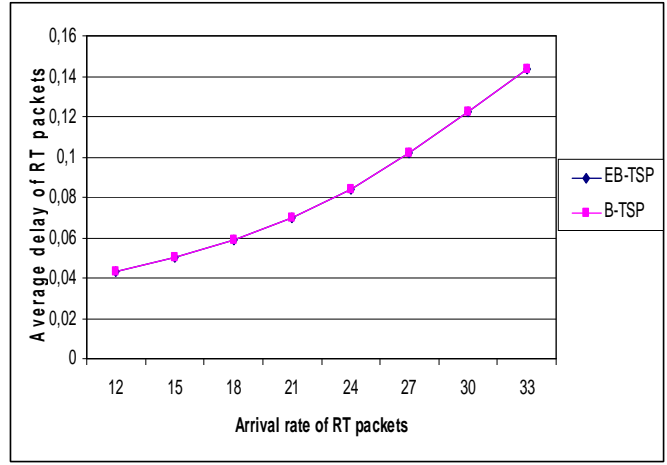

*Figure 3: Variation of the average delay of RT packets according to arrival rate of RT packets*

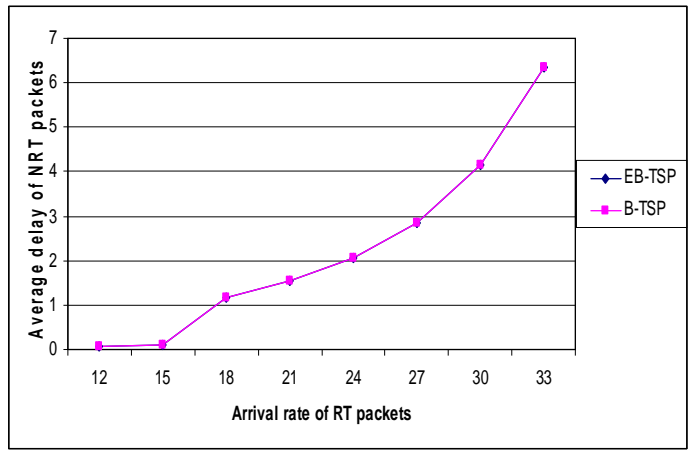

*Figure 4: Variation of the average delay of NRT packets according to arrival rate of RT packets*

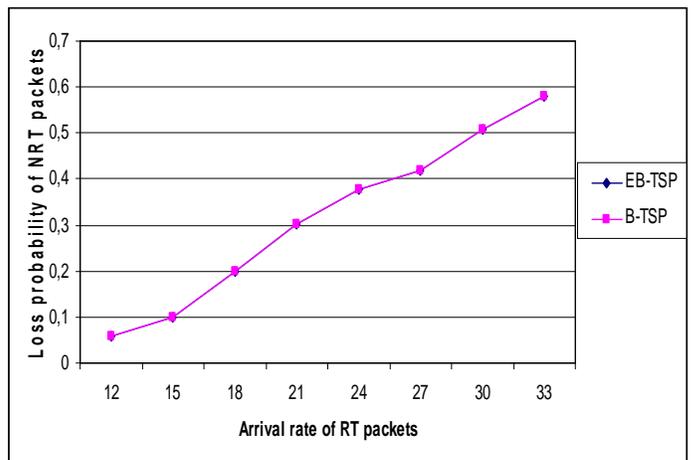

*Figure 5: Variation of the loss probability of NRT packets according to arrival rate of RT packets*

## VI. CONCLUSION

In this paper we have applied a new time space priority scheme (Enhanced Basic-TSP) in HSDPA where multiple flows exist for an end user. This scheme overcomes a limitation of the Basic-TSP scheme previously studied in the literature, and achieves a better management for buffer space.

We devise an ergodic continuous-time Markov chain CTMC to characterize the transition of the system. The QoS measures in the proposed scheme are analytically given for both flows. Numerical results show that the EB-TSP have a significant impact on the RT packet dropping, and keep the RT delay and NRT packet dropping in the same level compared to Basic-TSP scheme. This implies an enhancement of the QoS relatively to the received RT flow at the end users

### REFERENCES

[1] A.A. Abdul Rahman, K.Seman and K.Saadan, "Multiclass Scheduling Technique using Dual Threshold," APSITT, Sarawak, Malaysia, 2010.J. Clerk Maxwell, A Treatise on Electricity and Magnetism, 3rd ed., vol. 2. Oxford: Clarendon, 1892, pp.68–73.

[2] K. Al-Begain, A. Dudin, and V. Mushko, "Novel Queuing Model for Multimedia over Downlink in 3.5G", Wireless Networks Journal of Communications Software and Systems, vol. 2, No 2, June 2006.

[3] K. Al-Begain, Awan I. " A Generalised Analysis of Bffer Management in Heterogeneous Multi-service Mobile Networks", Proceedings of the UK Simulation Conference, Oxford, March 2004

[4] ] Choi, J. S. and C. K. Un, "Delay Performance of an Input Queueing Packet Switch with Two Priority Classes". Communications, IEE Proceedings- Vol.145 (3). 1998

[5] A. El Bouchti , A. Haqiq, M. Hanini and M. Elkamili "Access Control and Modeling of Heterogeneous Flow in 3.5G Mobile Network by using MMPP and Poisson processes", MICS'10, Rabat, Morocco, 2-4 November 2010.

[6] A. El bouchti and A. Haqiq "The performance evaluation of an access control of heterogeneous flows in a channel HSDPA", proceedings of CIRO'10, Marrakesh, Morocco, 24-27 May 2010.

[7] S. El Kafhali, M.Hanini, A. Haqiq, "Etude et comparaison des mécanismes de gestion des files d'attente dans les réseaux de télécommunication" . CoMTI'09, Tétouan, Maroc. 2009.

[8] Floyd, S and V. Jacobson.. "Random Early Detection Gateways for Congestion avoidance" , *IEEE/ACM Trans.Network*, Vol 1, No. 4. 1993

[9] Borko Furht and Syed A . Ahson, "HSDPA/HSUPA Handbook". CRC Press 2011.

[10] R. Nelson, "probability, stochastic process, and queueing theory", Spriger-Verlag, third printing, 2000.

[11] M. Hanini, A. Haqiq, A. Berqia, " Comparison of two Queue Management Mechanisms for Heterogeneous flow in a 3.5G Network", NGNS'10. Marrakesh, Morocco, 8-10, july, 2010.

[12] Pao, D. C. W. and S. P. Lam, "Cell Scheduling for Atm Switch with Two Priority Classes". ATM Workshop Proceedings, IEEE. 1998.

[13] G. Shabtai, I.Cidon and M.Sidi, "Two priority buffered multistage interconnection networks". Journal of High Speed Networks 15, IOS Press. 2006

[14] J.L. Van den Berg, R. Litjens and J. Laverman, "HSDPA flow level performance: the impact of key system and traffic aspects". MSWiM-04, Venice, Italy.2004.

[15] X.wang.H.Schulzrinne, " comparison of adaptive internet multimedia applications", IEICE Trans.commun, Vol E82-B no.6. 1999

[16] S.Y.Yerima and K. Al-Begain "Evaluating Active Buffer Management for HSDPA Multi-flow services using OPNET", 3rd Faculty of Advanced Technology Research Student Workshop, University of Glamorgan, March 2008.

[17] S.Y.Yerima and Khalid Al-Begain " Dynamic Buffer Management for Multimedia QoS in Beyond 3G Wireless Networks ", IAENG International Journal of Computer Science, 36:4, IJCS_36_4_14 ; (Advance online publication: 19 November 2009)

[18] S.Y.Yerima, K. Al-Begain, "Performance Modelling of a Queue Management Scheme with Rate Control for HSDPA" , The 8th Annual PostGraduate Symposium on The Convergence of Telecommunications, Networking and Broadcasting, Liverpool John, U.K. 28-29 June 2007.